\documentclass[epj,twocolumn,floatfix,final]{svjour}
\usepackage{graphicx}
\usepackage[usenames]{color}

\begin{document}
   
 
\title{Josephson oscillation of a superfluid Fermi gas}


\author{Sadhan K. Adhikari\thanks{Electronic
address: adhikari@ift.unesp.br; 
URL: http://www.ift.unesp.br/users/adhikari/}}

\institute{Instituto de F\'{\i}sica Te\'orica, UNESP $-$ 
 S\~ao Paulo State University,
 01.405-900 S\~ao Paulo, S\~ao Paulo, Brazil}

\date{\today}


\abstract{Using the complete numerical solution of a time-dependent
{{\it three-dimensional}} mean-field model we 
study the Josephson 
oscillation of
a superfluid Fermi gas (SFG) at zero temperature formed in a combined
axially-symmetric harmonic plus one-dimensional periodic optical-lattice
(OL) potentials after displacing the harmonic trap along the axial OL
axis.  We study the dependence of Josephson frequency on the strength of
the OL potential.  The Josephson frequency decreases with increasing
strength as found in the experiment of Cataliotti {\it et al.}  [Science
{ 293} (2001) 843] for a Bose-Einstein condensate {and 
of the experiment 
of Pezz\`e {\it et al.} [Phys. Rev. Lett. 93 (2004) 120401] for an 
ideal Fermi gas.} We demonstrate a
breakdown of Josephson oscillation in the SFG for a large displacement
of the harmonic trap.  These features of Josephson oscillation of a SFG
can be tested experimentally.} 

\PACS{{03.75.Ss} {Degenerate Fermi gases},
{03.75.Lm} {Tunneling, Josephson effect, Bose-Einstein condensates in 
periodic potentials, solitons, vortices, and topological excitations}, 
 {03.75.Kk} {Dynamic properties of condensates; collective and 
hydrodynamic excitations, superfluid flow}}
 

 \authorrunning{S. K. Adhikari}
\titlerunning{Josephson oscillation of a superfluid Fermi gas}

\maketitle

\section{Introduction}
 
The observation of an oscillating Josephson current across
an one-dimensional (1D) periodic array of potential wells, usually 
generated by
a
polarized 
standing-wave laser field and commonly known as an optical-lattice
(OL) potential, in a trapped ``cigar-shaped" Bose-Einstein condensate 
(BEC) 
by
Cataliotti {\it et al.} \cite{cata} was the first manifestation of this
phenomenon in trapped neutral bosons.  Until then the Josephson effect
was confirmed in superconductors with charged electrons and in liquid
helium \cite{cata,3}. 
In the experiment of Cataliotti {\it et al.} \cite{cata} a Josephson
oscillation was initiated in a repulsive $^{87}$Rb BEC formed in a
1D periodic OL plus an axially-symmetric harmonic
potentials by
suddenly displacing the harmonic trap along the axial direction. They
found that
the Josephson frequency reduced with
the increase of the strength of the OL  potential. Also a breakdown of 
Josephson oscillation was observed for large displacement of the OL 
potential. 

There have been theoretical studies of Josephson oscillation 
using the numerical
solution of the time-dependent mean-field Gross-Pitaevskii (GP) equation
\cite{8} in 
one (1D) \cite{cata,cata2} and three (3D) 
\cite{sad1,x} dimensions to understand different experimental  features 
 \cite{cata}. 
There have  also been other  theoretical studies of  Josephson oscillation 
in a trapped BEC \cite{str,str1} using different  approaches and
under different conditions distinct from the experiment of 
Cataliotti {\it et al.} \cite{cata}. Moreover, there have been many 
interesting studies on different aspects of Josephson oscillation in a 
cold Fermi gas under diverse  conditions of trapping \cite{ferjos}.

{Recently, in  another experiment on ideal Fermi gas of 
$^{40}$K atoms in 
an axially-symmetric plus OL trap  
Pezz\`e {\it et al.} \cite{pez}
observed the oscillation of the Fermi gas  after a sudden displacement 
of its equilibrium position along the lattice for various OL strength. 
They measured the frequency of oscillation for various OL strength.
They also explained the 
result of experiment using a semiclassical theory. 
}

Due to a strong repulsive Pauli-blocking interaction at low energies among
spin-polarized fermions, there cannot be an evaporative cooling leading to
a quantum degenerate Fermi gas (DFG) \cite{exp1}.  Trapped DFG has been
achieved only by sympathetic cooling in the presence of a second boson or
fermion component.  Recently, there have been successful observation
\cite{exp1,exp2,exp3,exp4} and associated experimental
\cite{exp5,exp5x,exp6} and theoretical \cite{yyy1,yyy,zzz,capu,ska}
studies of BEC-DFG mixtures. 
Different experimental
groups \cite{exp1,exp2,exp3,exp4} observed the formation of a DFG
in the following systems: $^{6,7}$Li
\cite{exp3}, $^{23}$Na-$^6$Li \cite{exp4} and $^{87}$Rb-$^{40}$K
\cite{exp5,exp5x}.  Also, there have been studies of a degenerate mixture
of two hyperfine-spin 
components of fermionic $^{40}$K \cite{exp1} and $^6$Li \cite{exp2}
atoms. Later on, the observation  \cite{Jin} of the  
transition of a DFG to a Bardeen-Cooper-Schrieffer (BCS) 
superfluid fermion gas (SFG)  of opposite hyperfine-spin orientation 
by manipulating the fermion-fermion interaction using a 
Feshbach resonance  has opened the possibility of controlled study of a 
BCS superfluid formed under the action of a weak fermion-fermion 
attraction. The use of a Feshbach resonance has also allowed the 
observation of BCS-BOSE crossover \cite{cross}  in 
two-hyperfine-component fermion 
vapors of  
$^{40}$K \cite{BCSJin,Jin}   and 
$^6$Li  \cite{grimm}  atoms.

There have been theoretical models of a DFG using a mean-field 
hydrodynamic approch 
based essentially on
the  Thomas-Fermi-Weizs\"acker approximation \cite{jz,capu}. Such models 
have 
provided results \cite{ska}
for collapse in a 
boson-fermion mixture of $^{87}$Rb-$^{40}$K
in qualitative agreement with the observation by 
Modugno {\it et al.}
\cite{exp5,zzz}.   
There have also been  studies of  fermionic bright \cite{BS}, dark 
\cite{DS},   
and gap  solitons \cite{EPL} in a
boson-fermion
mixture  
where  the bosonic component is treated by
the mean-field GP equation \cite{8} and the fermionic
component  by a hydrodynamic model \cite{ska,capu}. 
(The Feshbach resonances in the 
boson-fermion systems $^{23}$Na-$^6$Li and
$^{87}$Rb-$^{40}$K have been observed experimentally
\cite{feshf} and can be used in a controlled experiment of solitons in 
such a mixture.) 
The results of such 
studies are in agreement with {\it ab initio} studies of gap 
\cite{salerno}
and bright \cite{kar} solitons 
based on properly 
antisymmetrized many-body approach.

Although a mean-field-hydrodynamical approach to a DFG has its 
limitations 
(as in the absence of a coherent phase it may determine only the fermion 
density), 
a Ginzburg-Landau-type mean-field 
description \cite{fw}
of a BCS superfluid based on a Lagrangian density
is theoretically well founded and widely 
appreciated and has been used \cite{EPL,sala} in dealing 
with superfluid boson-fermion 
mixtures 
(such a description for a SFG determines the fermion density 
as well as the 
phase of a complex
coherent order parameter). A rigorous antisymmetrized many-body approach 
for a SFG becomes unfeasible as the number of fermions 
increases, where the simplified mean-field approach of a BCS 
superfluid in terms of a one-body wave function of Cooper-paired 
fermions 
is     of advantage.

In this paper we study the Josephson oscillation of a BCS
superfluid at zero temperature 
in an OL  potential using the complete numerical solution of 
a  3D
mean-field model.  As in the experiment on Josephson oscillation of a 
BEC \cite{exp5}, we consider a SFG formed in a 
cigar-shaped axially-symmetric parabolic 
trap with an added OL potential along the axial direction.  The 
Josephson oscillation is initiated by giving a translation of the 
parabolic trap in the axial direction upon the formation of the SFG. The 
SFG acquires energy in the process to initiate the Josephson 
oscillation. We study the variation of the frequency of Josephson
oscillation for different OL strength. We also study the breakdown of 
smooth oscillation for large initial displacement of the parabolic trap. 
{We compare our results for frequency with experimental 
results  on 
superfluid Bose $^{87}$Rb atoms \cite{cata} as well as on Fermi $^{40}$K 
atoms \cite{pez}. }

The present model is derived as the Euler-Lagrange equation of a 
Lagrangian density using the well-known energy density of a SFG   
\cite{Yang3D}. The Euler-Lagrange equation so obtained is 
a nonlinear partial differential equation with a nonlinearity of power 
7/3. Such an equation 
has also been used for a DFG with a different coefficient 
multiplying the nonlinearity (the 
coefficient in the two cases is different due to the Cooper pairing in a 
BCS superfluid).

In Sec.  II  we present the mean-field  model for a
SFG mixture. In Sec. III we present  numerical results for
Josephson oscillation of a SFG in a
combined harmonic plus 1D OL
potentials as a function of the strength of the OL  
 when the harmonic trap is displaced through a small distance 
along the optical axis. We also
illustrate the breakdown of  Josephson oscillation  in a SFG 
when the oscillation is initiated by giving a large displacement of the
harmonic trap along the optical axis.
 Finally, in Sec.  IV we present a summary  of our study.

\section{Mean-field Model for Fermi superfluid}

Let us consider a BCS  superfluid Fermi gas (SFG) of $N$ Cooper-paired 
fermions of mass 
$m$. 
The energy density of the system is given by ${\cal E}=3 \sigma 
\epsilon_F/5$ \cite{Yang3D}, where $\epsilon_F\equiv \hbar^2 k_F^2/(2m)$ 
is the 
Fermi energy with $k_F$ the Fermi momentum  and 
$\sigma$  the atomic density. (Here we neglect small corrections to this 
expression due to the residual fermion-fermion interaction usually 
expressed as an expansion in scattering length.)  
Modification to this expression appropriate 
to study a BCS-Bose crossover \cite{cross} has also been suggested 
\cite{salasnich}.  
The atomic 
density for the superfluid corresponding to the isotropic distribution 
(spherical Fermi
surface) is $\sigma={2(2\pi )^{-3}}\int_{0}^{k_{F}}4\pi
k^{2}dk\equiv \left( 3\pi ^{2}\right) ^{-1}\left( 2m\varepsilon 
_{F}/\hbar
^{2}\right) ^{3/2}$, where the effect of Cooper 
pairing is included in the extra factor of 2. This relation can be 
inverted to yield $\epsilon_F=\hbar^2 (3 \pi^2\sigma)^{2/3}/(2m)$, which 
leads to the following expression for energy density of the system
\begin{equation}
{\cal E} = \frac{3\hbar^2(3\pi^2)^{2/3}}{10m}\sigma^{5/3}.
\end{equation}

In the spirit of the Ginzburg-Landau theory \cite{fw} the SFG  can be 
described by a complex order parameter $\Psi$, such that 
$\sigma=\Psi^2$.
In terms of this order parameter, the Lagrangian density of the 
superfluid can be written as 
\begin{eqnarray}\label{lden} 
{\cal L}&=& \frac{i\hbar}{2}\left(\Psi^*\frac{\partial \Psi}{\partial t}
-\Psi\frac{\partial \Psi^*}{\partial t} 
\right)-\frac{\hbar^2}{2m_{\mathrm {eff}}}|\nabla \Psi|^2-V({\bf r})\Psi 
\nonumber \\
&-&
\frac{3\hbar^2(3\pi^2)^{2/3}}{10m}\Psi^{10/3},
\end{eqnarray} 
where $V({\bf r})$ is the external potential, and $m_{\mathrm {eff}}$ is 
the effective mass of superfluid flow as in the Ginzburg-Landau theory.
(The exact value of $m_{\mathrm {eff}}$ is not known but there is 
evidence \cite{fw} that for a 
Cooper-paired SFG  it is  twice the fermion 
mass and we shall use this value in 
the following: $m_{\mathrm {eff}}=2m$).
The Euler-Lagrange equation  of  Lagrangian density 
(\ref{lden})  becomes  the following  3D
nonlinear Schr\"odinger equation with a repulsive nonlinear term
of power $7/3$:
\begin{eqnarray}
 i\hbar \Psi _{t} = \frac{-\hbar ^{2}}{2m_{\mathrm{eff}}}\nabla^2 
\Psi
+\frac{\hbar ^{2}}{2m} \left( 3\pi ^{2}\right) ^{2/3}|\Psi
|^{4/3} \Psi+ V(\mathbf{r})\Psi ,  
\label{threeD1}
\end{eqnarray}%
with normalization $
\int |\Psi (\mathbf{r},t)|^{2}d{\bf r}=N.  $
In the presence of the
combined axially-symmetric and periodic
OL  potentials \cite{cata,sad1}
     $  V({\bf
r}) =\frac{1}{2}m_{\mathrm{eff}} \omega ^2(\rho ^2+\nu^2 z^2) 
+V_{\mbox{opt}}$, where
 $\omega$ is the angular frequency of the harmonic potential 
in the radial direction $\rho$,
$\nu \omega$ that in  the
axial direction $z$, with $\nu$ the aspect ratio, 
and $V_{\mbox{opt}}= s E_R\cos^2 (k_Lz)$ is
the OL potential  
created with the standing-wave laser field of wavelength 
$\lambda$,  
with $E_R=\hbar^2k_L^2/(2m)$, $k_L=2\pi/\lambda$, and $s$ $ (<12)$
the 
strength. Although, for small $s$ one should have a Joshepson 
oscillation, for large $s$ values Joshepson
oscillation should terminate, as in the case of a BEC \cite{mo}, due to 
a 
superfluid to Mott-insulator transition. (This transition cannot be 
studied with the mean-field equations  and field-theoretic analysis is 
needed for its understanding.)

In the axially-symmetric configuration, in the zero angular momentum 
state 
the fermion order parameter  
can be written as 
${\Psi({\bf r}, t)}= \psi(\rho,z,t)$, where $0\le  
\rho < \infty$  and $-\infty <z<\infty $.
Now  transforming to
dimensionless variables $\hat \rho =\sqrt 2 \rho/l$,  
$\hat z=\sqrt 2 z/l$,   $\tau
=t
\omega, $
$l\equiv \sqrt {\hbar/(m_{\mathrm{eff}}\omega)}$,
and
${ \varphi(\hat \rho,\hat z;\tau)} \equiv   
\hat \rho \sqrt{{l^3}/(N{\sqrt
8})}\psi(\rho,z;t),$  Eq.  (\ref{threeD1}) becomes \cite{x,ska}
\begin{eqnarray}\label{d1}
&\biggr[&-i\frac{\partial
}{\partial \tau} -\frac{\partial^2}{\partial
\hat \rho ^2}+\frac{1}{\hat \rho }\frac{\partial}{\partial \hat \rho}
-\frac{\partial^2}{\partial
\hat z^2}
+\frac{1}{4}\left(\hat \rho ^2+\nu^2 \hat z^2\right)  -{1\over \hat \rho
^2} 
\nonumber \\
&+& \frac{V_{\mbox{opt}}}{\hbar \omega}+
 n\left|\frac {\varphi({\hat \rho ,\hat z};\tau)}{\hat
\rho}\right|^{4/3}
 \biggr]\varphi({ \hat \rho,\hat z};\tau)=0,
\end{eqnarray}
where the nonlinearity parameter 
$ n=m_{\mathrm{eff}}(3\pi^2N)^{2/3}/m$. In terms of the 
1D  probability 
 $P(z,t) \equiv 2\pi\- \- \int_0 ^\infty 
d\hat \rho |\varphi(\hat \rho,\hat z,\tau)|^2/\hat \rho $, the
normalization of the
wave 
function 
is given by $\int_{-\infty}^\infty d\hat z P(z,t) = 1.$

In many of the experiments with DFG \cite{exp1,exp5,exp5x}, $^{40}$K
fermion atoms were
used and
in this study we also consider  $^{40}$K atoms
 and take $m$ to be the mass of K atoms.  As in the experiment of 
Cataliotti {\it et al.} \cite{cata}
the axial and radial trap frequencies are taken as  $\nu \omega =
2\pi \times 9 $ Hz and $ \omega =
2\pi \times 92$ Hz, respectively, with $\nu = 9/92\approx 0.1$. 
For $^{40}$K, the
harmonic-oscillator length $l=\sqrt {\hbar/(m_{\mathrm{eff}}\omega)} 
\approx 1$
$\mu$m and
the
present 
dimensionless length unit  corresponds to $l/\sqrt 2 \approx 0.7$
$\mu$m. 
The
present
dimensionless time unit corresponds to $\omega ^{-1} =
1/(2\pi\times 92)$ s $=1.73$ ms. Although we perform the calculation in
dimensionless units using Eq. (\ref{d1}), we present the results in
actual physical units using these conversion factors consistent with 
$^{40}$K atoms. We take the wavelength $\lambda$ 
of the standing-wave laser beam to make the 
OL potential to be  $\lambda=700$ nm, so that the          
dimensionless laser wave
length $\lambda _0= \sqrt2\lambda/l \simeq 1$ and  the dimensionless 
standing-wave energy parameter $E_R/(\hbar \omega)= 4\pi^2/\lambda _0^2$.
Hence in 
dimensionless units 
the  OL potential  of 
Eq.   (\ref{d1}) is
\begin{equation}\label{pot}
\frac{ V_{\mbox{opt}}}{\hbar
\omega}\equiv s \frac{E_R}{\hbar\omega}\cos^2(k_Lz)
=s \frac{4\pi^2}{\lambda_0^2} 
\left[
\cos^2 \left(
\frac{2\pi}{\lambda_0}\hat z
\right)
 \right].
\end{equation}
{Though most of our calculation was done with above 
set of parameters, 
for a comparison with the experiment on Fermi $^{40}$K 
atoms of 
Pezz\`e {\it et al.} \cite{pez}, we also repeated our calculation with 
trap frequencies  $\nu \omega =
2\pi \times 24 $ Hz and $ \omega =
2\pi \times 275$ Hz, and OL wavelength $\lambda =863$ nm.}

We solve  the {three-dimensional} Eq.  
(\ref{d1}) numerically  using a   
split-step time-iteration
method
with  the Crank-Nicholson discretization scheme described recently
\cite{11}.  
We discretize the GP equation typically with time step 0.001
and space step 0.1 spanning $\rho$ from 0 to 7 $\mu$m  and $z$ from 
$-120$
$\mu$m to
120 $\mu$m, although, sometimes we used smaller steps for obtaining
convergence. Equation  (\ref{d1}) is then solved by time iteration  
starting with the known harmonic oscillator solution for 
 $n=0$: $\varphi(\hat \rho, \hat z) = [\nu
/(8\pi^3)  ]^{1/4}$
$\hat \rho$ $e^{-(\hat \rho^2+\nu \hat z  ^2)/4}$ with chemical potential
$
\mu=(1+\nu /2)$
\cite{9}. For a typical cigar-shaped 
geometry with $\nu \simeq 0.1$
\cite{cata}, $\mu (\simeq 1)$ is much smaller than the 
typical depth of the 
OL  potential wells $E_R/(\hbar \omega) = 
4\pi^2 /\lambda
_0^2 \simeq
40$  so that $\mu
<<
E_R/(\hbar \omega)$ and the  passage of fermion atoms from one well to
other can only
proceed through quantum tunneling \cite{sad1,x}. 
During the time iteration of Eq. (\ref{d1})
the
nonlinearity  $n$ as well as the OL potential parameter
$s$ 
are  slowly increased by equal amounts in $100000$ steps of 
time iteration until the desired value of nonlinearity and OL
potentials are  attained. Then, without changing any
parameter, the solution so obtained is iterated 50000 times so that a
stable
solution  is obtained 
independent of the initial input
and time and space steps. 
The
solution then corresponds to the bound SFG under the joint action of
the harmonic and OL potentials.

\begin{figure}
\begin{center}
\includegraphics[width=.8\linewidth]{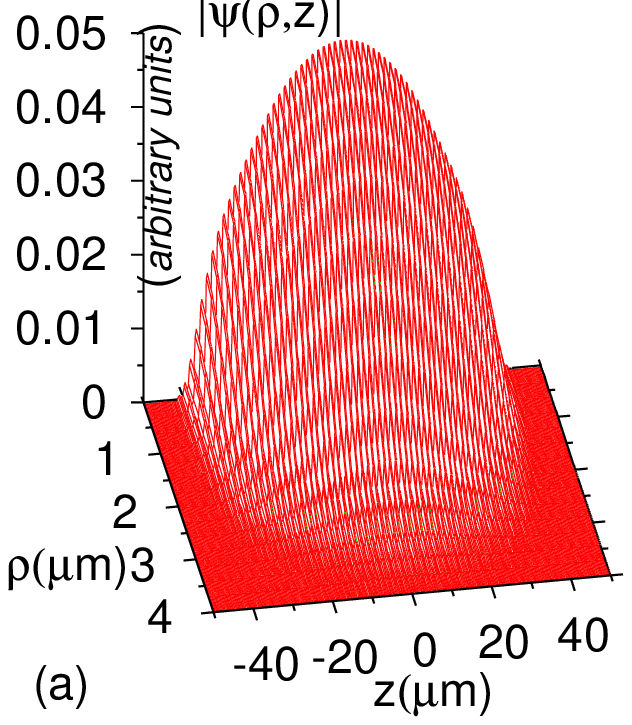}
\includegraphics[width=.8\linewidth]{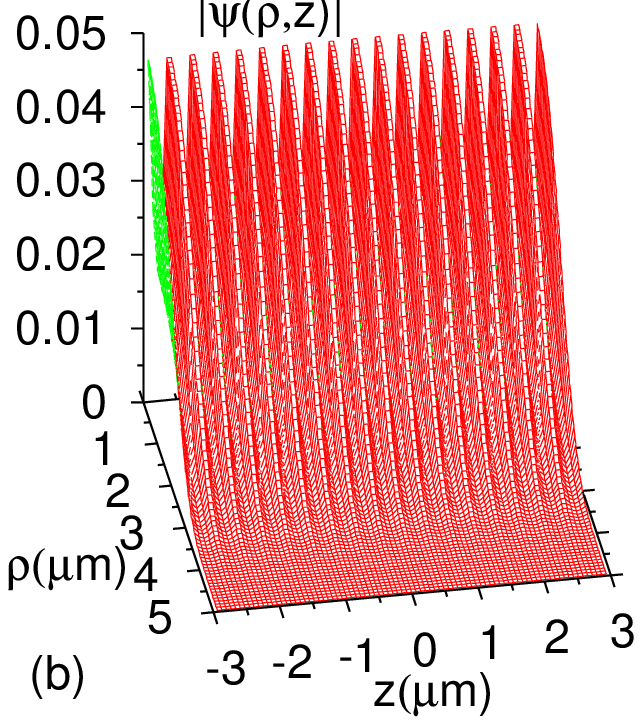}
\end{center}

\caption{(a) The profile of the SFG function $|\psi(\rho,z)|$
vs. $\rho$ and $z$ formed
in
the joint
optical-lattice and harmonic potentials for $N=40$, $n=225$ and $ 
s=4$.
(b) The profile of the  SFG function $|\psi(\rho,z)|$    of (a) 
 vs. $\rho$ and $z$  for 3 $\mu$m  $>z>-3$ $\mu$m,  which clearly
exhibits the maxima 
and minima of the SFG  function along the axial direction.}

\end{figure}

\section{Numerical Results} 

First we consider a SFG  formed in the combined harmonic and
periodic OL potentials for a specific 
nonlinearity $n$. 
We study the formation of a SFG in the combined harmonic and
OL potentials of  Eq. (\ref{pot}) for a range
of values of $s$.  
In Fig. 1 (a) the plot of $|\psi(\rho,z)|$ vs. $\rho$ and $z$
illustrates the profile of the SFG  for $N=40$, $n=225$ and
$s=4$. From Fig. 1 (a) we see that the SFG has the shape of a cigar (of 
length 80 $\mu$m and 
transverse radius 3 $\mu$m)
  cut into narrow slices with the
OL barriers separating the slices. 
The large number of maxima
and minima due to the OL potential is not 
clearly visible in
this plot.  
The maxima and minima in the axial direction are clear in the
plot of  $|\psi(\rho,z)|$ vs. $\rho$ and $z$            
in
Fig. 1 (b) where we show the central part of  function
$|\psi(\rho,z)|$   for 3
$\mu$m $>z >-3$ $\mu$m. 
In this interval of $z$, there are about 16
wells of the OL  potential and as many maxima and 
minima in $|\psi(\rho,z)|$.

\begin{figure}
 
\begin{center}
\includegraphics[width=.75\linewidth]{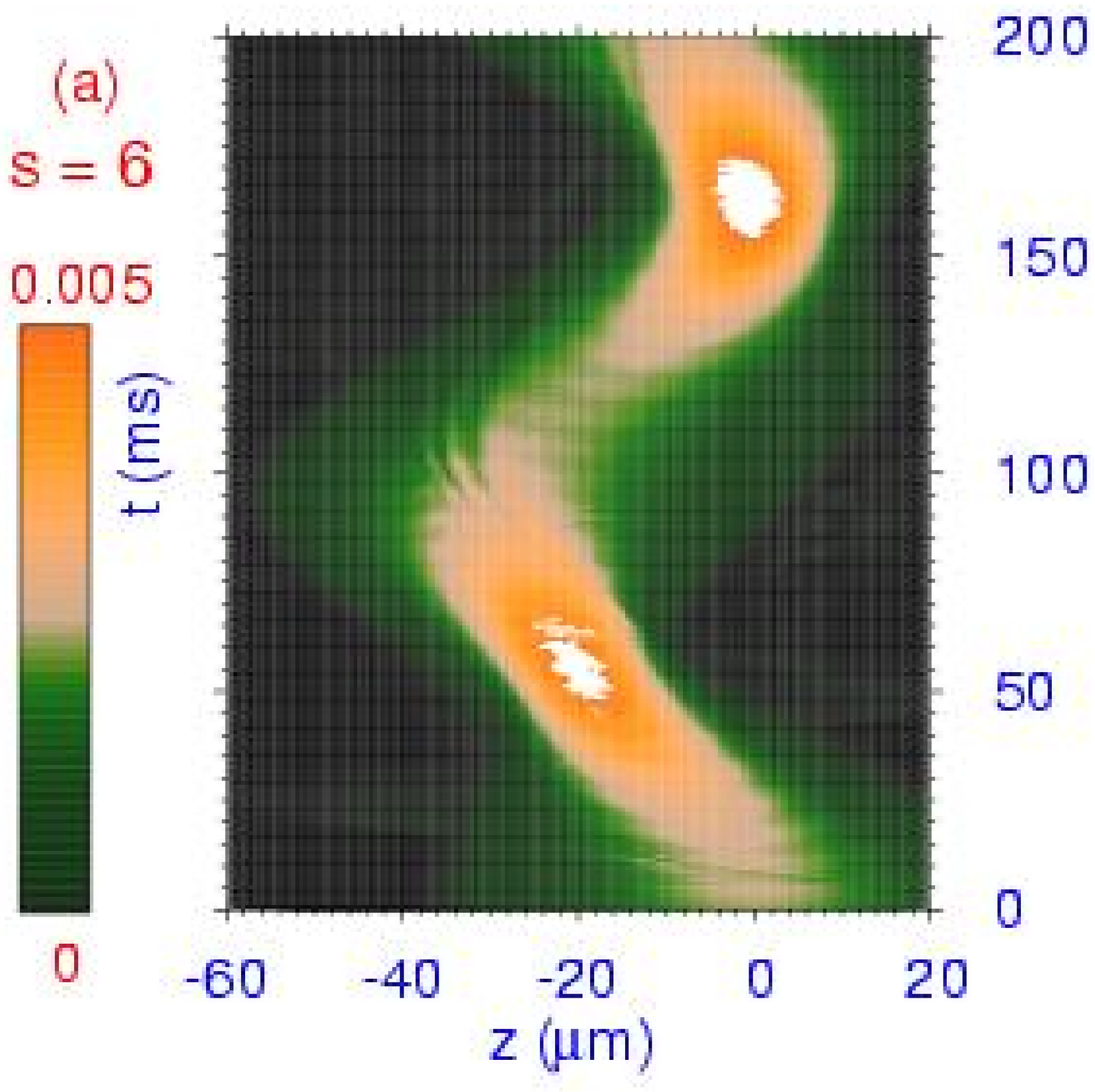}
\includegraphics[width=.75\linewidth]{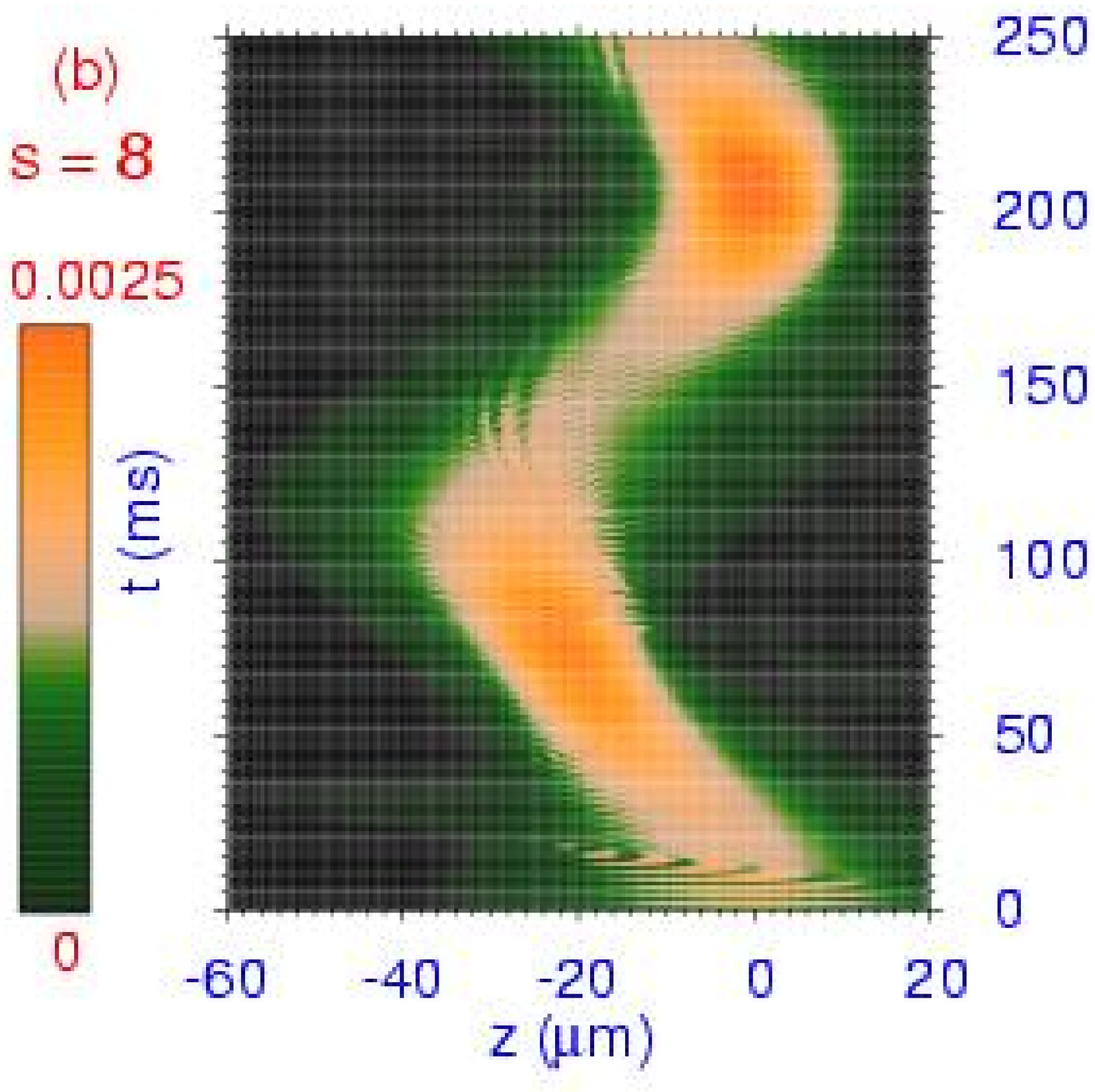}
\end{center}

\caption{(Color online) The contour plot of the 1D
probability density $P(z,t)$ of the SFG with $n=225$ executing a
Josephson oscillation when the harmonic trap is suddenly displaced along
the axial direction through a distance 14 $\mu$m at $t=0$ for
OL strength $s =$ (a) 6 and  (b) 8. The period of Josephson
oscillation can be obtained from these plots. }
 
\end{figure}

Now we consider an oscillating SFG in the combined harmonic and
periodic
OL potentials. If
we suddenly displace the harmonic trap along the lattice axis by
a small distance after the formation of the SFG in the combined
potentials, the
SFG will acquire a potential energy, 
be out of equilibrium and start to oscillate. As the
height of the potential-well barriers of the OL potential is
much
larger than the chemical potential of the system, the atoms in the SFG
will
move by tunneling through the potential barriers. This fluctuating
transfer of  atoms across the potential barriers is due to Josephson
effect in a neutral quantum SFG.

\begin{figure}
 
\begin{center}
\includegraphics[width=1.\linewidth]{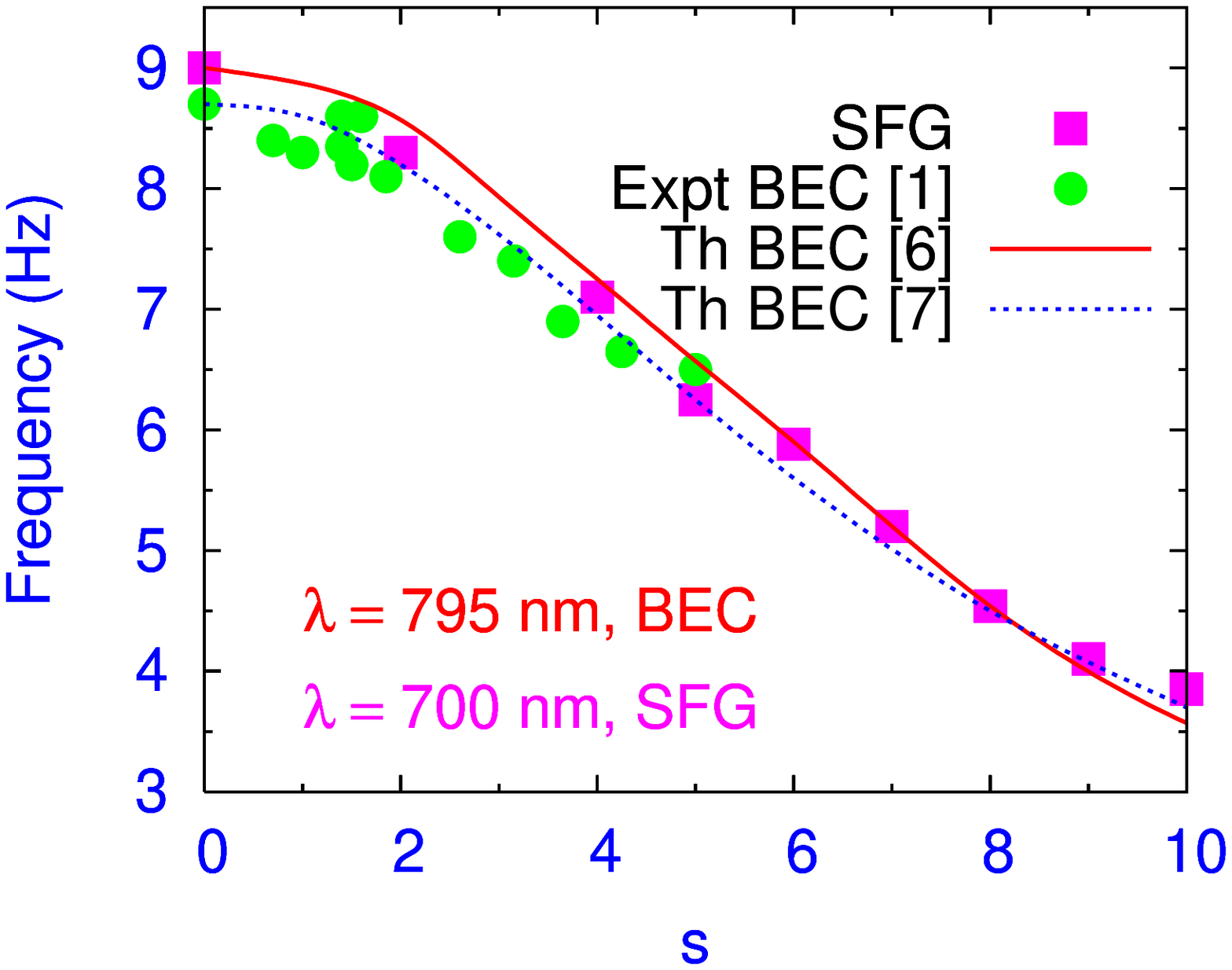}
\includegraphics[width=1.\linewidth]{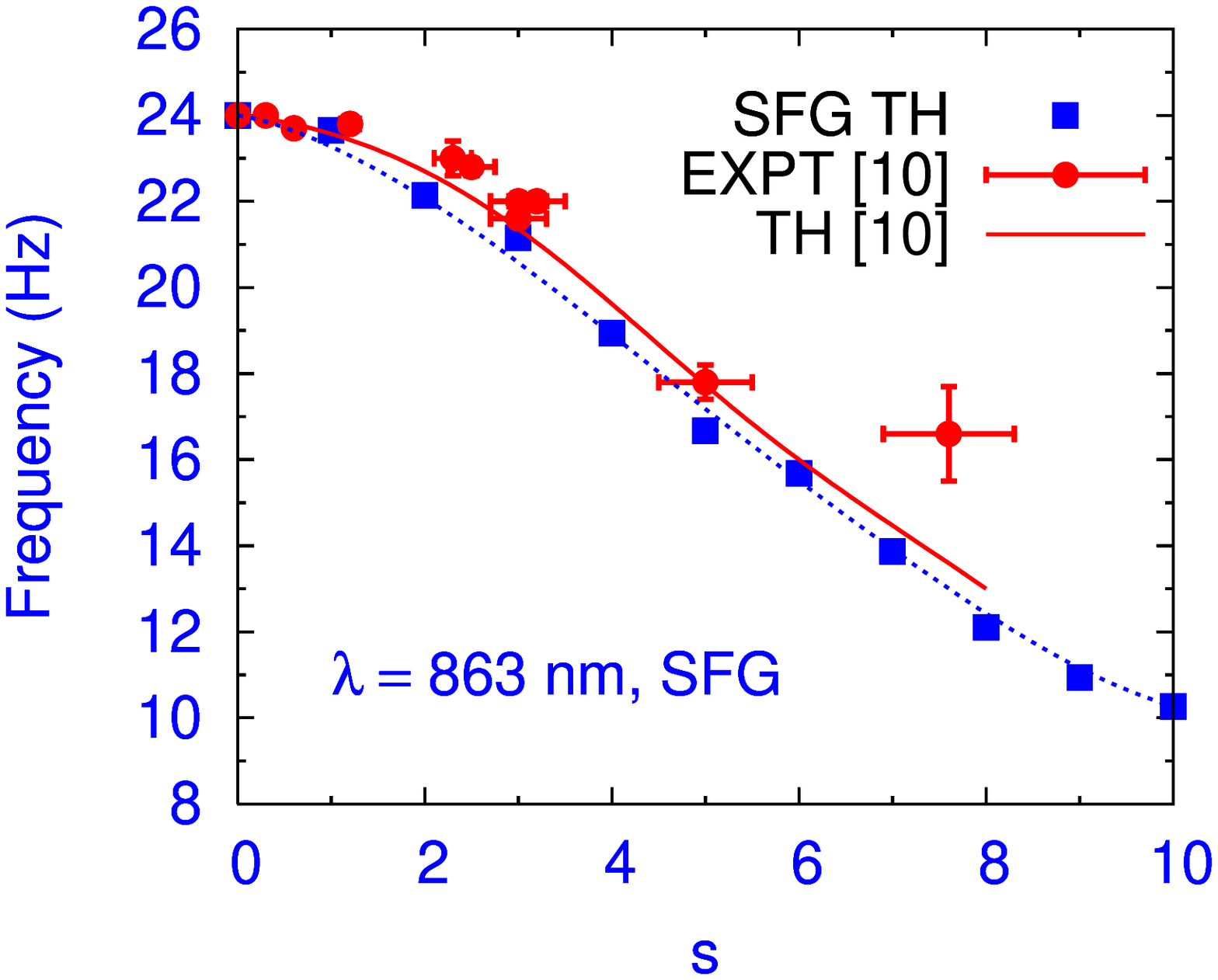}
\end{center}

\caption{(Color online) The frequency of the atomic current in the array
of Josephson
junctions as a function of OL strength $s$   compared with (a) 
experiment on boson $^{87}$Rb and other theories (b) on fermion 
$^{40}$K and other theory. In (a) solid 
square  
$-$ present result for SFG;   
solid circle 
 $-$ experiment of Cataliotti {\it et al.} \cite{cata} for BEC with
repulsive $^{87}$Rb atoms; 
full line   $-$ mean-field simulation for BEC \cite{x}; 
dotted line $-$ hydrodynamical calculation for BEC
 \cite{str}.  In (b) solid
square
$-$ present result for SFG;  solid circle
 $-$ experiment of Pezz\`e {\it et al.} \cite{pez} for fermions with
 $^{40}$K atoms; 
full line $-$ semiclassical theory of Pezz\`e {\it et al.} \cite{pez} 
for fermions; dotted line $-$ joins present points to guide eye.
}

\end{figure}

With the SFG of Fig. 1 (a)  we next study its Josephson 
oscillation
when the harmonic potential is suddenly displaced along the axial
direction 
by 14 $\mu$m. The SFG now acquires an added potential energy which it 
can
use to execute  a Josephson oscillation along the axial direction. The
Josephson oscillation is best studied numerically from the contour plot of
the 1D  probability $P(z,t)$ vs $z$ and $t$ exhibited
in Figs. 2. These plots
clearly show the central position of the SFG along the axial $z$
direction. In the present simulation we take  different
values of the OL strength $s$. These contour plots 
are very useful to find the Josephson frequencies. From
Figs. 2 the periods of Josephson oscillation are easily read off and the
frequencies calculated for different OL strengths $s$. 

In Fig. 3 (a) we plot the Josephson frequencies vs. OL strength
$s$. Specifically, in addition to the
present calculation for SFG  we also  show (i) the
experimental frequencies  of
Cataliotti {\it et al.}   \cite{cata}  for a repulsive BEC of $^{87}$Rb 
atoms, (ii) the 
3D
simulation for a BEC  from Ref. \cite{x}, and (ii) the
hydrodynamical calculation for a BEC  from Ref. \cite{str}.
{In Fig. 3 (b) we show the Josephson frequencies vs. OL 
strength with the parameters of the experiment of Pezz\`e  {\it et al.} 
  \cite{pez} for an ideal Fermi gas of $^{40}$K atoms. In this figure we 
compare our results with experiment and 
a semiclassical theoretical calculation   \cite{pez}.  }

The most interesting conclusion from Figs. 3 is that the present
frequencies of
3D simulation for Josephson oscillation of a SFG 
 are  
practically the
same
as the frequencies for Josephson oscillation of a BEC 
{and close to 
those for oscillation of an ideal Fermi gas.} 
For $s=0$ the OL
potential is absent
and the BEC and the SFG  execute free oscillation with the frequency of 
the
axial potential.
From Figs. 3 we find that  the Josephson
frequency reduces with increasing OL strength.  As
the Josephson oscillation takes place by quantum tunneling of the atoms
through the OL barriers, this oscillation is bound
to be reduced as the height of the OL barriers is 
increased. The reduction of Josephson oscillation with the increase of the
parameter $s$ results in a reduction of Josephson frequency in Fig. 3. 
{ The agreement of the frequencies of present model for 
superfluid 
Fermi gas with the experiment on ideal (nonsuperfluid) Fermi gas is 
remarkable. A future experiment on a superfluid Fermi gas may reveal 
subtle differences in behavior, if any,   of such oscillations in 
superfluid and 
nonsuperfluid 
Fermi gas.}

\begin{figure}
 
\begin{center}
\includegraphics[width=.48\linewidth]{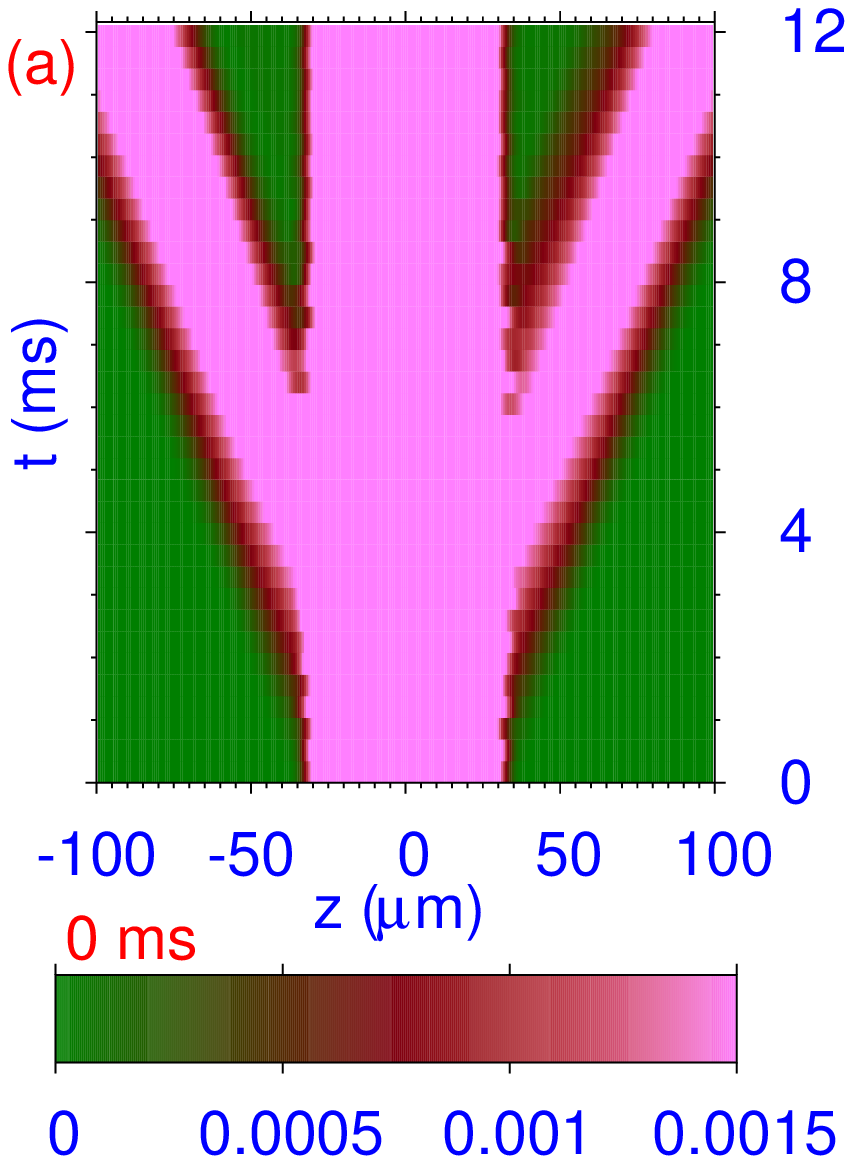}
\includegraphics[width=.48\linewidth]{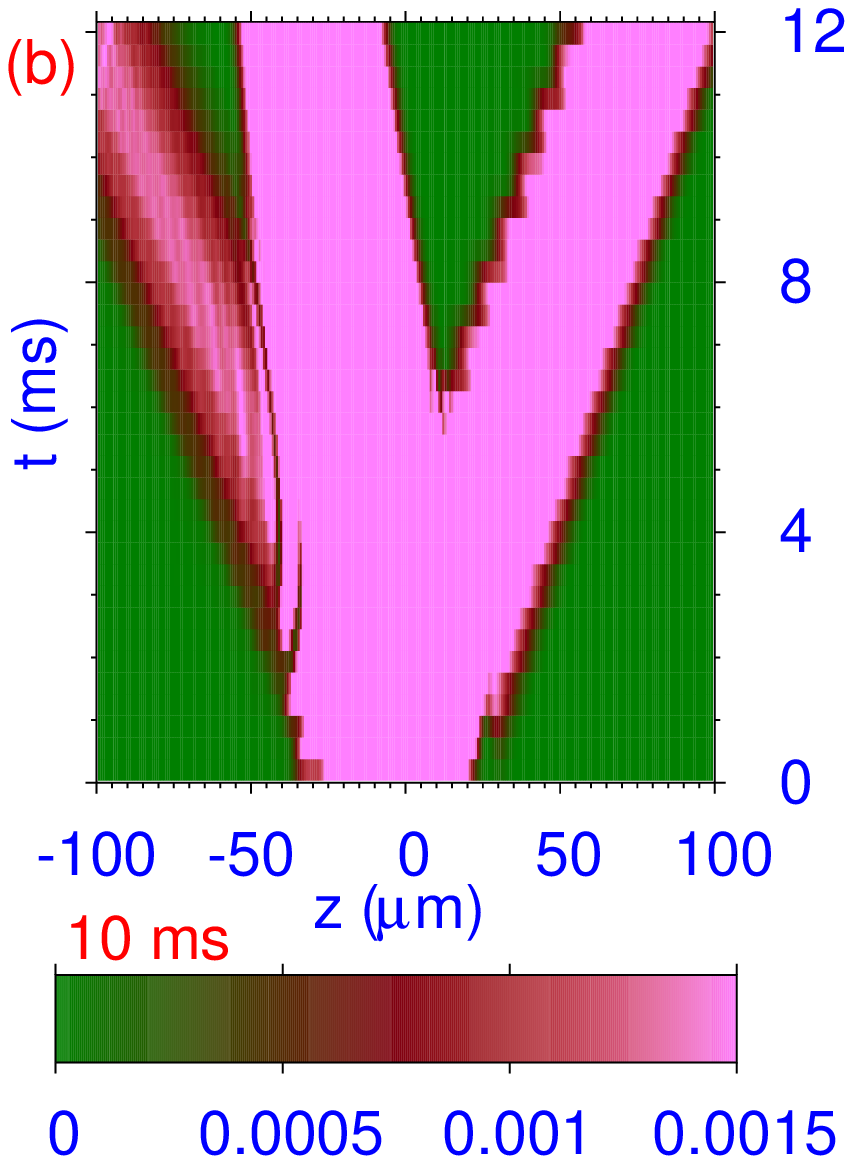}
\includegraphics[width=.48\linewidth]{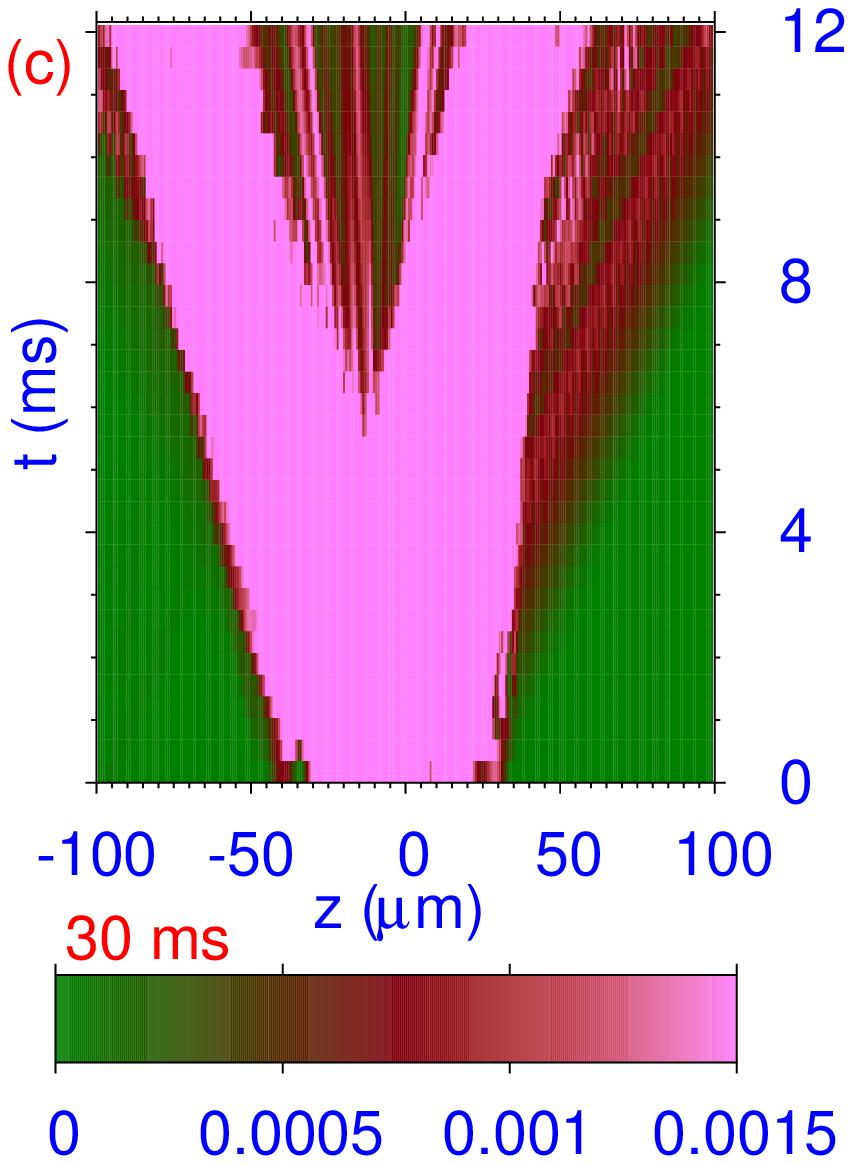}
\includegraphics[width=.48\linewidth]{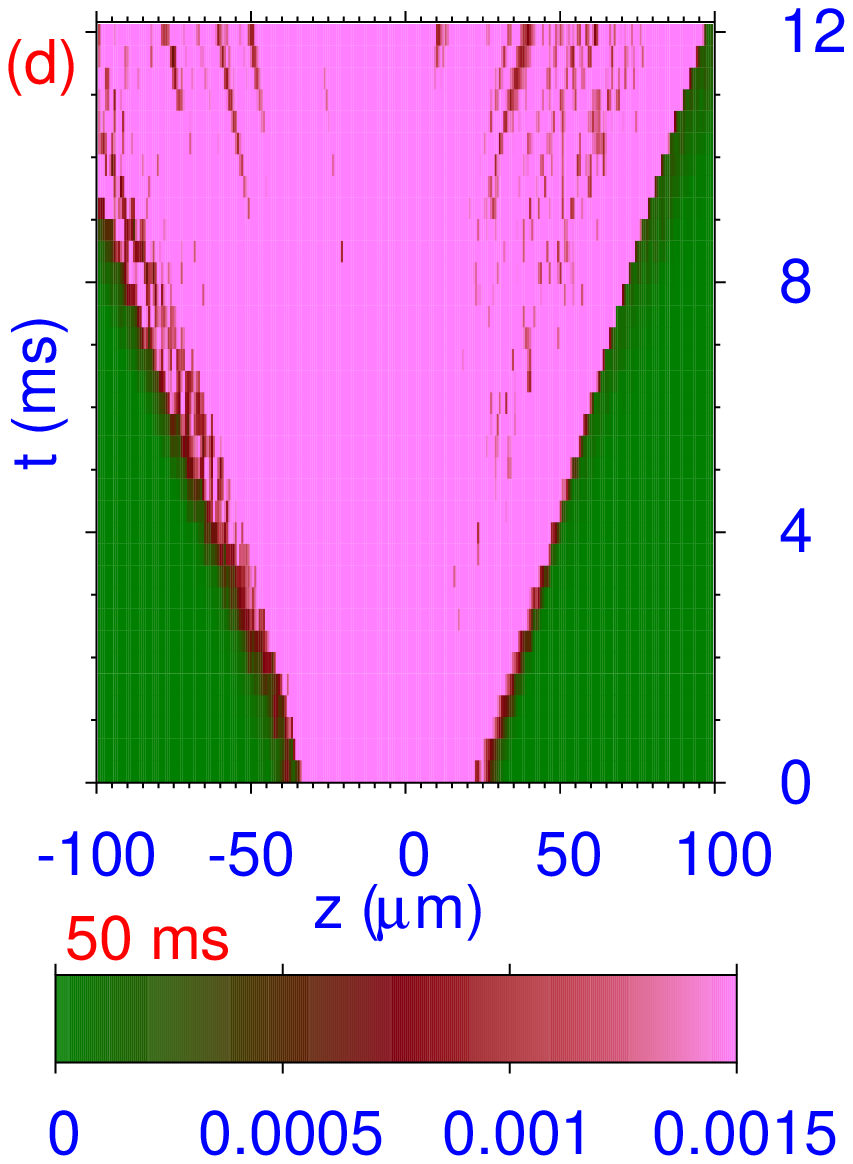}
\end{center}

\caption{(Color online) The contour plot of the 1D
probability density $P(z,t)$ of the SFG with $n=225$ released after 
hold times $T_h=$ (a) 0 ms, (b) 10 ms, (c) 30 ms, and (d) 50 ms
when the harmonic trap is suddenly displaced along
the axial direction through a distance 100 $\mu$m   for
OL strength $s = 8$. The three clean interference trails in
(a)  demonstrates the phase correlation in the initial SFG. With the
increase of hold times in the displaced trap the clean interference trail 
is slowly lost in (b) and (c) signalling a loss in phase correlation. The
phase correlation in the SFG is completely lost in (d) after a hold time
of
50 ms.}

\end{figure}

The SFG formed on the combined OL and harmonic traps considered so
far in Figs. 1 and 2 is a phase-correlated
quantum fluid with  the
atoms  freely moving by quantum tunneling from one OL site to
another at zero temperature and  one has a phase-coherent macroscopic 
state. 
This
will lead to an
interference pattern when the SFG is released from the OL
trap. However, the details of the pattern  may change from one experiment
to another. 
When the harmonic trap is displaced through a small distance as
in Figs. 2, the  SFG executes Josephson oscillation by
quantum tunneling.

Next we consider an  account  of the breakdown of Josephson
oscillation
in a SFG  when the harmonic potential 
is 
displaced by a very large distance ($\sim $ 100 $\mu$m) along the
periodic
OL  potential. 
In that case for small times ($t<300$ ms) no Josephson oscillation of the
type observed in Figs. 2 was found. The SFG was found to remain
virtually fixed at
$z=0$ for small times. At very large times it moved very slowly to the new
center of equilibrium. However, the phase correlation of the SFG is lost
after staying a significant time in the displaced trap. 
To demonstrate the destruction of phase correlation  in the displaced trap
we rely on the disappearance of the 
interference pattern formed upon the release of the SFG
from the combined harmonic and OL traps after a hold time of 
$T_h$ in the displaced trap. 

The formation of a BEC on a combined harmonic plus  OL
potentials 
and the interference pattern formed upon the release of this BEC
from the confining traps have been studied both theoretically
\cite{cata2,muru,muru1}
and
experimentally \cite{cata,mo,mo1}. We present a similar study for the 
SFG.
The periodic OL potential cuts  the SFG into several pieces
at different sites maintaining phase correlation among them. As a result
when the SFG is released from the traps it expands
freely and a matter-wave interference pattern is formed in a few
milliseconds. The atom cloud released from one lattice site expands, 
overlaps and interferes with atom clouds from neighboring sites to form a
robust interference pattern, consisting of a central peak and two smaller
symmetrically spaced peaks moving in opposite directions
along the OL axis \cite{cata,mo,mo1}. Since the
lattice transfers momentum to the SFG in units of $2p_R=2h/\lambda$, the
recoil velocity of each the two side peaks is given by $v_R=
2p_R/(2m)=2h/(2m\lambda)$ \cite{cata,muru1}. Using $l^2=
\hbar/(2m\omega)$, we
have $v_R=
4\pi l^2 \omega /\lambda\approx 10 $ mm/s, where we used the numerical
values $l\approx 1$ $\mu$m, $\omega^{-1}=1.73$ ms, and
$\lambda\approx 1/\sqrt2$ $\mu$m.

To demonstrate the loss of phase correlation after displacing the harmonic
trap along the OL direction through 100 $\mu$m, in Figs. 4 we
exhibit the contour plot of the 1D probability density
$P(z,t)$
of the SFG with $n=225$ released after different hold times $T_h$ in 
the
harmonic plus OL traps. In Fig. 4 (a) we exhibit the result
of simulation for $T_h=0$, which has no effect on the phase
correlation. Upon release from the traps a robust interference pattern of
the central and two side peaks can be seen in this figure. The side peaks
move with velocities 10 mm/s and in 12 ms they move about 120 $\mu$m each
as can be seen from this figure. The successive figures 4 (b), (c), and
(d) show the situation for hold times 10 ms, 30 ms, and 50 ms,
respectively. The clean interference pattern of a central and two
symmetrically moving side peaks is destroyed slowly with the increase of
hold time.  For an intermediate $T_h$ (= 10 ms and  30 ms) there is
partial
destruction of the interference pattern.
For $T_h=50$ ms, there is no interference pattern and the 
expanding SFG occupies the full region between the two side peaks.
This simulation illustrates that when the harmonic trap is displaced by a
large distance, there is no Josephson oscillation.

The breakdown of Josephson oscillation above is
due to a classical transition from a phase-correlated 
superfluid to an ``insulator" resulting in a modulational instability as 
in a BEC \cite{bd}.
Other mechanisms for the loss of phase coherence in a BEC have also
been studied \cite{phco}.  The loss of phase coherence considered in all
these investigations \cite{bd,phco} originated from a classical
superfluid to insulator transition, different from a quantum transition
of a superfluid to a Mott insulator observed in Ref. \cite{mo}.
The classical phase transition involves energy, whereas the quantum phase
transition does not involve a supply of energy.

{A recent investigation \cite{ano} indicates that, for a 
BEC, the disruption of the large amplitude oscillations is caused by the 
onset of dynamical instability occurring when the quasi-momentum 
surpasses a critical value towards the edge of the first Brilluoin zone.
In this regard, it would have been interesting to study what is the 
critical quasi-momentum as a function of the nonlinearity of a SFG.   
In another experimental study on superfluid Fermi gas \cite{ano2}
with an OL moving 
at a constant speed, the   
breakdown of 
superfluidity was measured as function of speed over the entire 
Bose to BCS  crossover. A careful theoretical investigation on these 
issues, although beyond the scope of present work, 
are welcome   in the future.}

\section{Summary and Conclusion}

We  performed 3D
numerical simulation based on a time-dependent 
mean-field equation  to study  Josephson oscillation of a BCS 
superfluid
in
a combined axially-symmetric harmonic and periodic
OL  potentials. The OL potential is aligned along the axial direction 
and is
created by a standing-wave laser beam. The Josephson oscillation is
initiated by displacing the harmonic potential along the OL
axis by a small distance ($\sim 15$ $\mu$m). We study the variation of
Josephson frequency with the strength of
the OL potential and find that the frequency decreases with
the increasing OL strength $s$.  
As $s$ increases from 1 to 10 the frequency of our SFG
model is in agreement 
with an experiemnt on BEC and on ideal Fermi gas. 
However, for a large displacement ($\sim 100$ $\mu$m) of the harmonic
potential along the OL
axis we demonstrate a breakdown of Josephson oscillation. The SFG then 
returns very slowly to the new mean position without executing Josephson
oscillation. The breakdown of Josephson oscillation  also leads to a 
destruction of  
phase correlation and consequent loss of superfluidity 
in the Fermi gas  and 
is demonstrated by allowing the SFG to undergo a free
expansion. The absence of a distinct interference pattern upon free
expansion signals the loss of phase correlation  due
to a  classical phase
transition  to an insulating state as opposed to a quantum 
superfluid to a Mott  insulator transition observed in a BEC \cite{mo}.
Similar superfluid to insulator   classical phase
transition has been observed \cite{cata,bd}  and 
studied \cite{sad1,phco} in a BEC.   
 These features of
Josephson oscillation could be verified  experimentally 
and will provide a
test for the mean-field  model for a SFG.

A proper treatment of SFG should be performed using a fully
antisymmetrized many-body Slater determinant wave function
\cite{yyy1,kar} as in the case of atomic and molecular
scattering \cite{ps}. However, in
view of the success of the hydrodynamic model in other contexts
\cite{ska,BS,DS}
we do
not believe that the present study on Josephson oscillation of a DFG in 
an
OL potential to be so peculiar as to have no general validity.

\vskip 1cm

{I thank Dr. Paulsamy  Muruganandam for help in preparing the figures. 
The work was supported in part by the CNPq and FAPESP 
of Brazil.}


 \end{document}